# Ultrasonic tracking of a sinking ball in a vibrated dense granular suspension

S. Wildenberg[*], X. Jia[†], J. Léopoldès, and A. Tourin

Institut Langevin, ESPCI Paris, PSL University, CNRS, 1 rue Jussieu, 75005 Paris, France
[*] Currently at Laboratoire Magmas et Volcans, Université Clermont Auvergne, 63178 Aubière, France
[†] Corresponding author: xiaoping.jia@espci.fr

**Abstract**

Observing and understanding the movement of an intruder through opaque dense suspensions such as quicksand remains a practical and conceptual challenge. Here we use an ultrasonic probe to investigate the dynamics of a steel ball sinking in a 3D dense glass bead packing saturated by water. We show that the frictional model developed for dry granular media can be used to describe the ball motion induced by horizontal vibration. From this rheology we infer the static friction coefficient and effective viscosity that decrease when increasing the vibration intensity. Our main finding is that the vibration-induced reduction of the yield stress and increase of the sinking depth are presumably due to induced slipping at the grain contacts but without visible plastic rearrangements of grains, in contrast to dry granular packings. To explain these results, we propose a mechanism of acoustic lubrication that reduces the inter-particle friction and leads to a decrease of the yield stress. This scenario is different from the mechanism of liquefaction usually invoked in loosely packed quicksands where the vibration-induced compaction increases the pore pressure and decreases the confining pressure on the solid skeleton, thus reducing the granular resistance to external load.

**Introduction**

A ball dropped in a Newtonian fluid of smaller density starts to accelerate under gravity until reaching a terminal velocity, which for low Reynolds numbers is inversely proportional to the viscosity according to Stokes' law. This experiment is the basis of a classical method for measuring the viscosity of a Newtonian fluid. In yield stress fluids such as foams, emulsions and fluid-particle suspensions, sinking may arrest at a certain depth due to the jammed state (or yield stress) as function of the packing density, confining pressure and shear stress applied by the intruder[1-7]. Dense suspensions of non-colloidal particles occur in many industrial applications and geological processes such as mudflows and underwater avalanches[3,4]. Their flow dynamics is very rich and depends to a large extent on the density mismatch between the suspending liquid and the solid particles[8]. Density matched, or neutrally-buoyant suspensions, behave basically as non-Newtonian liquids without yield stress (except in the presence of the confining pressure between grains in contact[9]), while non-density-matched suspensions subject to gravity show roughly two flow regimes[8] similar to that of dry granular media[10,11]: a slow creeping flow in which the contacts between the particles are essentially frictional, and a fast inertial flow where the particles lose enduring contact and exhibit a Bagnold rheology due to particle inertia and brief collisional contact.

     More precisely, in a dry granular packing, the flow dynamics has been described by a modified frictional rheology $\mu(I) = \tau/P$ with $\tau$ the shear stress and $P$ the confining pressure. $I$ is the inertial number, i.e. the ratio between the characteristic time of grain rearrangement controlled by confining pressure and the macroscopic time determined by shear rate [11]. Such a concept was recently applied to dense granular suspensions in which the inertial relaxation time is rather governed by viscous drag forces at small Stokes number[12,13]. By using a viscous number $I_v$, a similar constitutive friction law as used in dry granular media has been proposed[8,9,13]. Aside from solid-like friction, there is another important characteristic that sets granular flows apart from classical Newtonian flows, namely non-locality[14,15] where the rheological response depends on the sample size; rheological studies performed in split-bottom and Couette cells provided evidence that flow in any part of a granular system leads to yielding throughout the entire system[16,17]. Similarly, studies in dry granular media[18-25] and gravitational granular suspensions[26,27] show that vibrations can also reduce the yield stress or/and modify the effective viscosity of flow.

However, in the case of natural phenomena such as liquefaction of quicksand[28, 29], several important questions still remain. Are there other mechanisms of liquefaction than vibration-induced compaction[4, 28], given that a heavy intruder can also sink in sheared dry granular packings[16]? Can the frictional rheology describe the motion of an intruder in vibrated dense granular media without plastic rearrangement of grains[30, 31]? Studies of an intruder moving inside dense granular media are mostly limited to 2D packings[32] and investigations within realistic 3D opaque granular media still pose challenging problems. Here we report on a new experimental approach to investigate the dynamics of a steel ball sinking in granular sediments, i.e. glass bead packs saturated by water (i.e. non-density-matched "suspensions"). We show that applied horizontal vibrations induce visible plastic rearrangements of grains in dry granular packings, but not in dense granular suspensions. Then, using a non-intrusive acoustic probe, we precisely monitor the position of the steel ball during its descent in the vibrated 3D opaque granular suspension. We analyse these results within the framework of previous granular rheological models[8, 9, 11-13] and we show that the transition from the solid state to the liquid state is intimately connected to the weakening of the shear resistance of contact via slip between grains[30, 33-35].

**Results**

Optical observation of the fine details of flow inside a 3D granular suspension is difficult due to light scattering by grains. One possibility to overcome this difficulty is to match the refractive index of the particles to that of the constituent fluid, but this technique is not applicable to most real opaque systems. In this work, we investigate the sinking of an intruder by tracking its position via its ultrasonic echoes as shown in Fig 1 (see Methods for details). Our set-up consists of a plexiglass container with diameter $Dc \approx 150$ mm filled with a suspension of glass beads of diameter $d \sim 100$ μm in water up to $H \approx 180$ mm. It is supported on a bearing (a rolling bead G) and coupled via a linker to a shaker, which applies a horizontal vibration at a frequency in the range of 50-300 Hz first for the sample preparation and then for the sinking experiment. The acceleration of vibration $a$, measured by the accelerometer A, ranges from $\Gamma = a/g = 0.27 - 2.7$ with $g = 9.8$ m$^2$/s. The horizontal vibration combined with gravitational sedimentation results in a dense granular suspension with initial packing density of about 62%. Interestingly, at this stage there is no visible rearrangement of glass beads under vibration in contrast to the case of dry packings (see movies 1 and 2 in Supplementary information) where the grains in the top layers flow or slide against each other whenever $\Gamma > \mu$ (i.e., $F_a/W > \mu$ with $F_a = ma$, $W = mg$) with $\mu$ the effective friction coefficient[20, 21]. The intruder is chosen from a set of steel balls of respective radii $R = 4, 5,$ and $7$ mm, carefully placed on the surface of this jammed granular packing. The ball begins to sink as soon as the horizontal vibration is applied to the container.

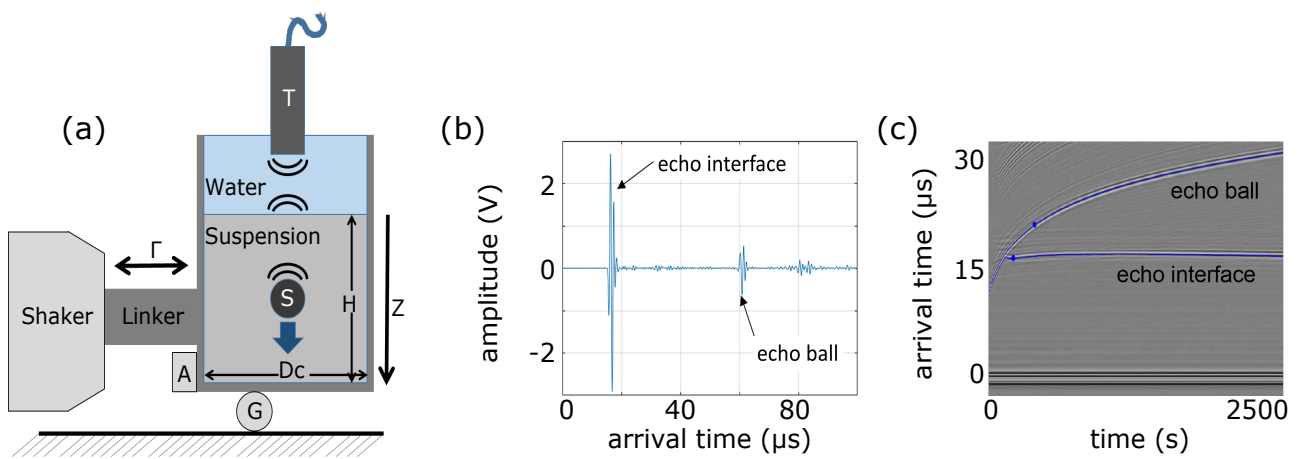

**Fig. 1.** (a) Sketch of the experimental set-up. (b) Ultrasonic echoes from the interface and from the intruder (ball). (c) Stack of waveforms showing the position of the intruder and the position of the sediment-water interface during a ball sinking experiment.

To characterize in detail the sinking of the intruder in the dense granular suspension, we first examine how the position $z(t)$ (Fig. 1a) depends on the acceleration of vibration $\Gamma$ and on the intruder diameter $2R$. Fig. 2a shows that the motion $z(t)$ of the intruder of radius $R = 5$ mm is smooth when $\Gamma$ varies from 0.27 to 2.7; we do not observe short periods of stalling as was reported for sinking in 2D dry granular media[32]. We then calculated, for each $\Gamma$, the instantaneous sinking velocity $dz/dt$ and the acceleration $d^2z/dt^2$ by numerical differentiation. More specifically, $dz/dt$ was obtained via $[z(t+\Delta t) - z(t)]/\Delta t$, where the time interval $\Delta t$ was set to 20 seconds as to limit noise while remaining much shorter than the duration of the data set. The acceleration $d^2z/dt^2$ was calculated from the velocity in a similar way. Figs. 2b and 2c depict the instantaneous velocity and acceleration of the ball as a function of the sinking position $z$, which shows up a fast deceleration at short times and a slow sinking at long times.

A cross-plot of the acceleration as a function of the velocity reveals a clear kink separating the fast decelerating and the quasi-steady flow regimes (Fig. 2d). The velocity at this kink $(dz/dt)^*$ is determined from the intersection of linear fits to the two regimes and we define $z^*$ and $t^*$ as the depth and time at which $(dz/dt)^*$ is reached. The depth $z^*$, slightly depending on $\Gamma$, is roughly equal to the diameter of the sinking ball; this suggests that the rapid decelerating regime occurs during the partial submergence of the intruder and is dominated by complex effects such as crater formation on the surface just above the intruder. In the following, we focus on the quasi-steady flow regime that is represented by the plots in Fig. 2e.

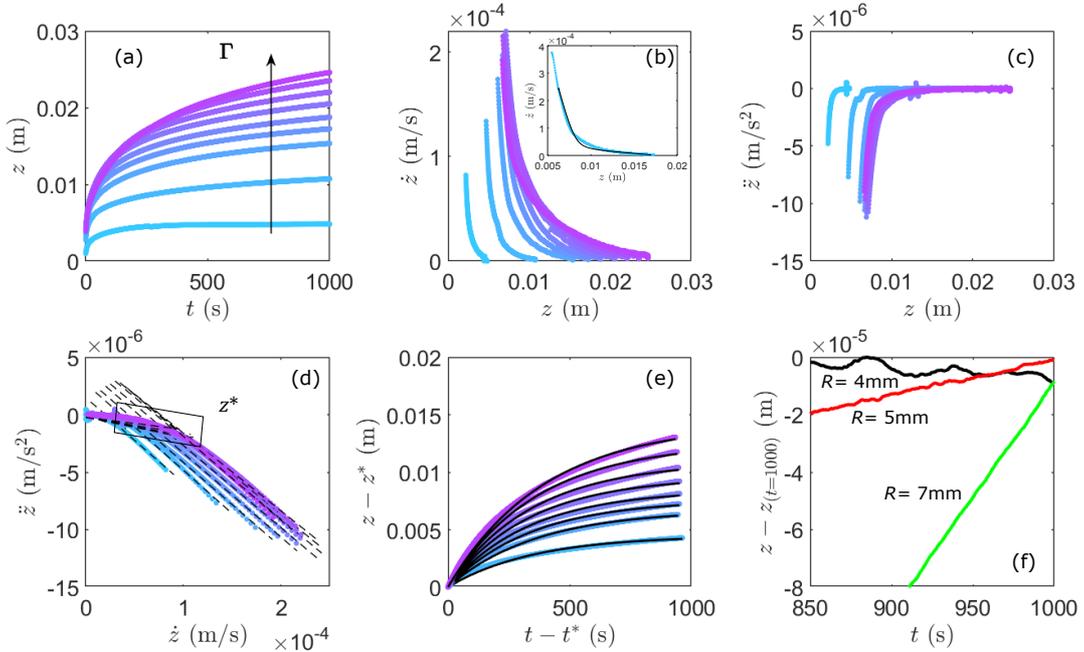

**Fig. 2.** Sinking dynamics of a steel ball of radius $R = 5$ mm for different vibration intensities $\Gamma$. (a) Positions of the ball versus time. (b) Instantaneous sinking velocities versus ball positions. Inset: comparison with the full analytical solution (black line) for $\Gamma = 1.08$ (see Eqs. 7&8 in Methods). (c) Accelerations versus ball positions. (d) Sinking acceleration versus velocity, which shows a transition from a fast decelerating regime to a quasi-steady regime. $(dz/dt)^*$ is obtained from the intersection of two sets of linear fits (see dotted lines). (e) The ball positions versus time in the quasi-steady regime. The black lines correspond to the solutions fitted in the quasi-steady regime with Eq. 4. (f) Evolution of the ball position at the lowest acceleration of vibration for various radii.

We observe that, for the lowest vibration $\Gamma$, the intruder decelerates until approaching a final depth at which it is partly submerged (Fig. 2a). We observe similar behaviour for the smaller (Fig. 3a1) and larger intruders (Fig. 3a2), respectively. As shown by zoomed traces in Fig. 2f, the small ball of radius $R = 4$ mm almost reaches the depth of arrest in the experimental time range and exhibits a fluctuating motion around it. Here the resolution for probing the relative displacement is $\Delta z = (1/2)\, c_w\, \Delta T_0 \sim 4$ μm provided by the sampling rate $1/\Delta T_0 \sim 200$ MHz. This fluctuation is reminiscent of the creep-like motion of a partially submerged intruder in a sheared dry granular medium[16]. For increasing values of $\Gamma$, the intruder asymptotically reaches larger depths (Fig. 2a): it keeps decelerating but does not completely stop within our measurement time. The influence of the intruder size is investigated by repeating the experiments with intruders of different size $R$ = 4 mm and 7 mm (Fig. 3). The phenomenology of sinking remains similar for all different intruders. Especially we recover the fast decelerating and the quasi-steady flow regimes. Interestingly, we find out that in the steady flow regime, the larger the intruder the larger the sinking depth.

Finally, we have tested whether or not the steel ball with $R$ = 5 mm will sink in a dry granular packing of glass beads ($d \sim 100$ μm) under the same vibration conditions. First, we do observe that the intruder starts sinking when the external vibration is applied; moreover it reaches the final depth more slowly than in the water-saturated granular packing (see movie 2 in Supplementary information). Second, we observe visible grain rearrangements (i.e. jiggling) at the surface of the dry granular packing in contrast to the vibrated dense granular suspension.

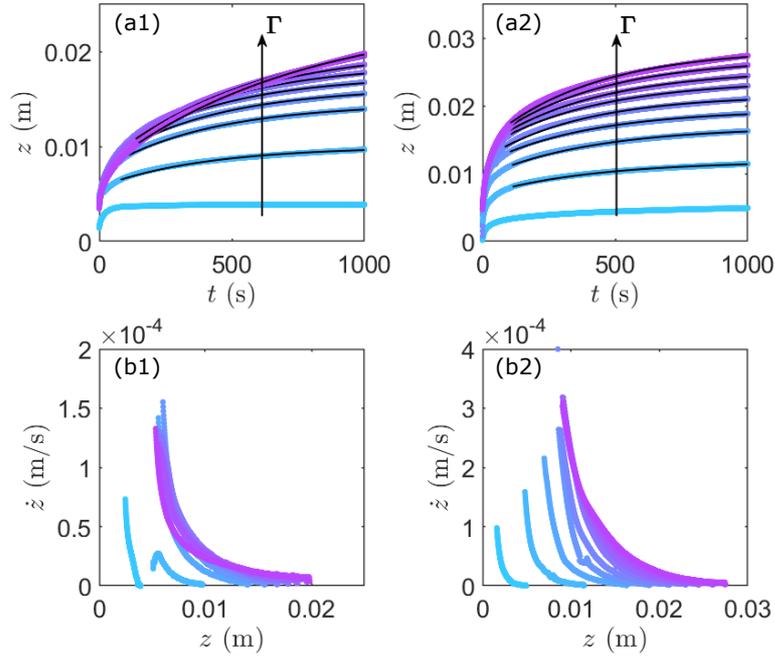

**Fig. 3.** Sinking dynamics for different intruder-sizes and for different vibration intensities $\Gamma$. (a1) Depth versus time for an intruder with $R$ = 4 mm and (a2) for an intruder with $R$ = 7 mm. (b1) Instantaneous velocity versus sinking depth obtained from (a1) and (b2) from (a2). The black lines correspond to the solutions fitted in the quasi-steady regime with Eq. 4.

**Discussion**

We consider the equation of motion for the intruder sinking along the vertical direction. There are three forces acting on the intruder during its trajectory through the suspension, a force due to weight, a buoyant force and a frictional force ($F_s$) due to the gravitational suspension (Fig. 4a):

$$m(d^2z/dt^2) = (4/3)\pi R^3 \Delta \rho_1 g - F_s \qquad (1)$$

with $m = (4/3)\pi R^3 \rho_s$ and $\Delta \rho_1 = \rho_s - \rho_{sus}$. Here $\rho_s \sim 8000$ kg/m³ is the mass density of the steel ball and $\rho_{sus}$ (= $0.62\rho_p + 0.38\rho_f$) ~ 1930 kg/m³ is the density of the granular suspension (mass densities of a glass bead and water are respectively $\rho_p$ = 2500 kg/m³ and $\rho_f$ = 1000 kg/m³). To specify the frictional force $F_s$, we need to know the frictional rheology defining the gravitational granular suspension. Courrech du Pont et al[12] and Cassar et al[13] have previously shown that the constitutive law of dense granular suspensions can be inferred from the relevant time scales controlling grain motion. In particular, one can evaluate the Stokes number St = $t_{av}/t_{ff} \sim 0.3$ for a confining pressure $P_g = \Delta \rho_2 g z$ due to gravity (with $\Delta \rho_2 = \rho_p - \rho_f$ and $z \sim 10$ mm the depth). Here $t_{av} = (2/3)\rho_p \alpha d^2/\eta_f$ is the characteristic acceleration time for a grain to reach a limit velocity $v_\infty = P_g \alpha d/\eta_f$ due to the viscous force and $t_{ff} = d(2\rho_p/(3P_g))^{1/2}$ is the time of free fall for a grain over a distance equal to one diameter $d$ without interstitial liquid (i.e., in a dry granular packing). $\eta_f/\alpha$ is the effective viscosity of the granular suspension with $\eta_f \sim 10^{-3}$ Pa.s the viscosity of water and $\alpha \sim 0.01$ a permeability parameter[13]. Besides, the particulate Reynolds number may be estimated by[6] $Re_p = \rho_p v_p d/(\eta_f/\alpha) \sim 10^{-4}$ with the flow velocity $v_p$ close to the sinking velocity of the intruder $dz/dt \sim 10^{-4}$ m/s (see discussion below). The small values of St < 1 and $Re_p \ll 1$ indicate that the interstitial fluid change the time scale of the microscopic rearrangements: the viscous forces dominate at the grain scale. A heuristic rheological law can then be written in terms of the dimensionless number $I_v$ [8, 12, 13]:

$$\tau = (\mu_0 + \mu_1 I_v)P_g = \mu_0 P_g + (\mu_1 \eta_f/\alpha)d\gamma/dt \qquad (2)$$

with $\tau$ the shear stress, $P_g$ the confining pressure, $d\gamma/dt$ the strain rate, $\mu_0$ and $\mu_1$ the static and dynamic (viscous) friction coefficients. The viscous number $I_v = t_{fall}(d\gamma/dt)$ is interpreted as the ratio between the time of grain rearrangement $t_{fall} = d/v_\infty = (\eta_f/P_g\alpha)$ and the characteristic time $(d\gamma/dt)^{-1}$ imposed by the shear. Equation 2 shows that for low shear rate, the local shear stress in a gravitational suspension is a combination of a (frictional) yield stress and a viscous stress, similar to the case of a Bingham plastic fluid with a yield stress $\tau_0 = \mu_0 P_g$ and a viscosity $\eta_B = \mu_1 \eta_f/\alpha$ [3, 6, 7].

There are few if any analytical solutions for a solid ball falling (or settling) through non-Newtonian suspensions. The flow field around the ball may differ appreciably from Stokes' solution in a Newtonian fluid due to the yield stress [Beris85]. Nevertheless, it was shown that for a ball sinking in Bingham plastic fluids at small Reynolds number, $Re_B = \rho_{sus}(dz/dt)^2/[\tau_0 + \eta_B(dz/dt)/2R]$ with $\gamma = (dz/dt)/2R$, the dependence of the drag coefficient $C_d = f(Re_B)$ can be tracked back to that of Newtonian fluids, i.e. $C_d \sim 24/Re_B$ [6]. Here it must be emphasized that the calculation of the Reynolds number includes the yield stress. Following the work of Dedegil[6], we consider the frictional force $F_s$ (Eq. 1) as the sum of two contributions, $F_s = F_0 + F_d$, with one threshold force $F_0 = (\pi R)^2(\mu_0 \Delta \rho_2 gz)$ independent of the Reynolds number $Re_B$ (< 0.1) and one drag force $F_d = C_d \pi R^2 \rho_{sus}(dz/dt)^2/2$. The existence of a yield stress $\tau_0 = \mu_0 \Delta \rho_2 gz$ leads to the fact that the intruder below a certain depth does not sink (or settle) at all but remains suspended. This is the case when the forces due to weight, buoyancy and threshold are balanced ($dz/dt = 0$) in Eq. 1, from which the depth of penetration is deduced to $z_\infty = (4/\pi)\Delta\rho_1 R/(3\mu_0 \Delta\rho_2)$ at a given $\mu_0$. This indicates that the larger the intruder $R$ or/and the lower the static friction coefficient the larger the penetration depth, as is illustrated by the asymptote of traces $z(t)$ in Figs. 2a, 3a1 and 3a2.

To interpret qualitatively the sinking dynamics, we use the above approximate expression for the drag coefficient in the laminar regime which leads to a Stokes-like viscous force $F_d = 6\pi R \eta_{eff}(dz/dt)$ with an effective viscosity $\eta_{eff} = \tau_0/(d\gamma/dt) + \eta_B$. Starting from this simple expression for the frictional force

$$F_s = (\pi R)^2(\mu_0 \Delta \rho_2 gz) + 6\pi R \eta_{eff}(dz/dt) \qquad (3)$$

and neglecting the slight variation of $\eta_{eff}$ in the low shear rate[27], we may derive the analytical solution of equation of motion (1) using the initial conditions (see Methods for details). In the quasi-steady regime $d^2z/dt^2 \to 0$, the corresponding solution with the initial condition $z = z^*$ at $t = t^*$ can be written as

$$z(t) = (z_\infty - z^*)[z_\infty/(z_\infty - z^*) - \exp{-k(t - t^*)}] \qquad (4)$$

with $k = (2/3)\mu_0 \Delta\rho_2 gR/\eta_{eff}$. This equation is used to fit the traces of $z(t)$ in the quasi steady flow regimes (black lines) in Figs. 2e, 3a1 and 3a2. The static friction coefficient $\mu_0$ and effective viscosity $\eta_{eff}$ obtained from such fitting for different diameters of the sinking ball are plotted as a function of the vibration acceleration $\Gamma$ in Figs. 4b and 4c, respectively. The corresponding values of $\mu_0 \sim 0.5$ and $\eta_{eff} \sim 10^4$ are consistent with those found in literature[8, 9, 27].

Similarly to previous studies[18-28], we consider the effect of horizontal vibration as a controlled perturbation, or effective temperature, that modifies the mechanical properties of granular media such as yield stress and effective viscosity. For the lowest $\Gamma$ the static friction coefficient deduced from this sinking ball experiment is about $\mu_0 \sim 0.7$, being comparable with the static friction coefficient $\mu_0^*$ (= tg$\theta_m$) $\sim 0.5$ that we measured from the avalanche angle $\theta_m \approx 25°$ (= tan$^{-1} \mu_0^*$) via inclined plane without vibration [9, 10]. When the vibration intensity $\Gamma$ is increased, $\mu_0$ and $\mu_1$ decrease rapidly until they approach a constant value for $\Gamma > 2$ g. Moreover, at low $\Gamma$, the values for $\mu_0$ and $\eta_{eff}$ appear to depend on the size of the sinking ball $R$; one possible explanation could be the failure of the assumption of local rheology (Eq. 2) on which the mean-field description is based. Indeed, the correlation length of force chains[4], i.e., 5-10$d$ (~ 0.5-1 mm) is not completely negligible in comparison with the intruder size $R$.

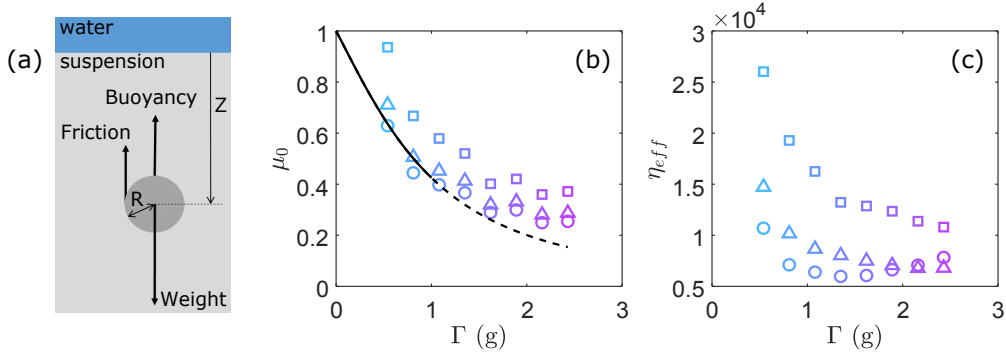

**Fig. 4.** (a) Sketch of a ball sinking in a dense granular suspension. (b) Static friction coefficient $\mu_0$ and (c) effective viscosity $\eta_{eff}$ versus vibration intensity, determined from the fit of traces $z(t)$ in Figs. 2a, 3a1 and 3a2 by Eq. 4. The different symbols correspond to the intruder sizes $R$ = 4 mm (circles), 5 mm (triangles) and 7 mm (squares). The dotted line in (b) corresponds to a heuristic friction model (see text).

Unlike rheological measurements previously reported in dry granular materials[23], we notice that the external vibration does not cancel the yield stress (or $\mu_0$ appearing in Eq. 2) in our granular suspensions for the range of vibration $\Gamma$ used here (Fig. 4b). As mentioned above[20, 21], when the horizontal vibration $\Gamma > \mu$ is applied to the cell containing a dry granular packing with free surface, the top layers of the material are liquefied and driven to the convective motion, while the low part moves with the container in solid body motion. This is precisely the depth at which the sinking ball gets jammed because the friction force becomes larger than the gravity (Eq. 1). Likewise, in dense granular suspensions, the friction coefficients $\mu_0$ and effective viscosity $\eta_{eff}$ (that we assumed to be spatially homogeneous in Eq. 2) can also be expected to vary in depth as the packing density. As described in Methods, we need to use an empirical depth-dependent effective viscosity $\eta_{eff}$ to reproduce the sinking dynamics at short times (inset of Fig. 2b). Numerical simulation would be useful to further investigate this issue[36].

Interestingly, here we do not observe any significant plastic rearrangement of grains nor convection on the surface of the horizontally vibrated dense granular suspension. Such striking difference between dry and liquid-saturated granular packings could be partly explained by the characteristic time of rearrangement it takes for a grain to move from a cage to the next one over a distance ~ $d$. For the dry packing we find that the free fall time is $t_{ff}$ [= $d(2\rho_p/(3P_g))^{1/2}$] ~ 3 ms, while for the granular suspension the fall time is $t_{fall}$ [= $\eta_f/(P_g\alpha)$] ~ 50 ms for a confining pressure $P_g$ ~ $d \rho_p g$ ~ 2.5 Pa. For the vibration frequency range ($f$ = 50-300 Hz) in our experiments, the period of the oscillatory driving is $T$ (= $1/f$) ~ 10 ms, which leads to $t_{ff}$ *(dry)* $< T < t_{fall}$ *(saturated)*. Under such conditions, we expect that the grains in the dry granular packing have the time to move or rearrange to the new cages before the applied vibration changes the driving direction. Instead the grains in the granular suspension have not the time to rearrange during one period of external vibration.

To understand such vibration-induced liquefaction in dense granular suspensions without plastic rearrangement, we propose a scenario that relies on the decrease of the friction coefficient $\mu_0$ (or yield stress) (Figs. 4b and 5) as follows. Since the acoustic wavelength in water at the vibration frequency $\lambda_w$ (= $c_w/f$ > 1 m) is much larger than the granular sample size, no gradient of acoustic pressure is induced; thus the saturating liquid should have a small impact on the elasticity of the solid skeleton (except buoyancy) in this surface-free granular suspension. Then, the weakening of the granular material is presumably ascribed to the modification of the contact network by the acoustic vibration. Indeed, it has been shown that the coefficient of friction between two smooth solid spheres can be reduced via a mechanism of acoustic lubrication[34], leading to the softening of the tangential contact stiffness and accordingly to a decrease of the sound velocities observed in granular packings[24, 30, 31]. To rationalize this hypothesis in disordered granular packing, we rely on a friction model developed for a multi-contact solid interface in which the shear interfacial stiffness is reduced through the slipping of asperities under oscillatory shear[33]. If we generalize this concept by replacing the asperities by the grain contacts, we can explain the softening of the shear modulus $G$ and thus of wave velocity observed in a 3D dense granular suspension[31] via the slipping of the contacts without any rearrangement of grains. As the yield stress $\tau_0$ (as well as $\mu_0 = \tau_0/P_g$) is roughly proportional to the shear modulus ($\mu_0 \sim G$) in a yield stress fluid[3, 27], we may adopt the formula developed for the shear modulus softening as[31, 33], $\mu_0/\mu_0^* \sim G/G_0 \approx 1/[1+\Gamma/(2\mu_p))+(5/4)(\Gamma/(2\mu_p))^2]$ induced by the external vibration $\Gamma$ (<1). Here $\mu_0^*$ is the static friction coefficient in dry granular media without external vibration and $\mu_p$ is the inter-particle friction coefficient. The dotted line in Fig. 4b represents the model prediction using $\mu_0^* \sim 1$ and $\mu_p \sim 0.7$ that compares qualitatively well with our data. These used friction coefficients are relatively high but compatible with values measured in yield experiments for granular media[4, 22] and for solid friction[33].

Finally we believe that this mechanism of acoustic lubrication may also partly explain the sinking of the intruder in the vibrated dry granular packing. But the effect of the packing density change needs to be accounted for due to the plastic rearrangements of grains[31]. Note that the vibration-induced random motions of grains in dry granular media observed here at the top surface were previously analysed[19, 22] using the concept of effective temperature determined by $\Gamma$. Interestingly, for a same external vibration $\Gamma$, our experiments showed that the ball takes a longer time in dry packings than in water-saturated packings to reach the final depth (see Supplementary Fig. S3). This result suggests that the effective viscosity (defined by Eq. 2 using the inertial number[13]) is higher in the dry case. Following this $T_{eff}$ concept, our observation would support the scenario expected by the fluctuation-dissipation relation[37] applied in granular media[19, 38]: at a given effective temperature $T_{eff}$, the larger the random force (i.e. grain collision) the larger the dissipation (i.e. effective viscosity).

In summary, we have developed a non-intrusive ultrasound method to monitor the dynamics of an intruder submerged in an opaque dense granular suspension. We have found that the sinking ball in the gravitational suspension under horizontal vibration reaches rapidly a quasi-steady regime, which is well described by the frictional granular rheology. The deduced static friction coefficients and effective viscosity decrease with increasing vibration intensity. The mechanism of acoustic lubrication that we proposed allows us to understand the vibration-induced fluidization, i.e. liquefaction in dense granular suspensions without plastic rearrangement of grains. Further studies are needed to investigate the creep-flow behaviour near jamming and the vibration-induced fluidization accompanied with grain rearrangements and collisions as in the case of dry granular packings. We believe that this work should help to get a better understanding of the liquefaction of quicksands as well as the triggering of landsliding.

**Methods**

*Ultrasonic tracking of an intruder.* The steel ball of radius $R$ is initially at rest on the surface of the densely packed granular suspension with solid volume fraction $\phi \approx 0.62$ measured as $m/(\rho_p V)$ where $m$ is the mass of glass beads, $\rho_p$ the glass density and $V$ the volume of the bead packing, respectively. To follow the ball sinking induced by vibration in our 3D opaque suspension we employ acoustic monitoring. For this purpose, a broadband transducer (denoted by T in Fig. 1a) centred at $f_{US}$ = 2.25 MHz is placed exactly above the intruder. It is used to emit short pulses and detect the echoes at a repetition rate ~ 100 Hz. The associated wavelength of ultrasound is $\lambda_{US} = c_w/f_{US} \approx 670$ μm with $c_w$ = 1500 m/s the sound velocity in water; it is smaller than the intruder but larger than glass beads ($d \ll \lambda_{US} \ll 2R$). In Fig. 1b we observe two ultrasonic echoes: one from the interface of the suspension and another one from the intruder. A stack of such waveforms (Fig. 1c) shows that the time of arrival of the echo from the interface remains constant, while the arrival time of the echo $T_0$ from the intruder increases as the distance between the transducer and the sinking intruder increases. We determine the position of the ball at time $t$ (during sinking) by $z(t) = (1/2) c_w T_0$.

*Analytical solution for a ball sinking.* For simplicity we rewrite Eq. 1 as,

$$d^2z/dt^2 = d - b(dz/dt) - cz \qquad (5)$$

with $d = \Delta\rho_l/\rho_s$, $b = (9/2)\eta_{eff}/R^2$, and $c = 3\mu_0\Delta\rho_2 g/(\rho_s R)$. The solution of this second order nonhomogeneous differential equation has the form:

$$z(t) = d/c + A \exp(\lambda_1 t) + B \exp(\lambda_2 t) \qquad (6)$$

where $\lambda_1 = (-b/2) + (b^2/4 - c)^{1/2}$ and $\lambda_2 = (-b/2) - (b^2/4 - c)^{1/2}$, $A$ and $B$ are the constants. Considering the initial conditions at $t_i$ ($\approx 14$ s, estimated from Fig. 2), the position of the ball is $z(t_i) = z_i \approx 0.006$ m and its velocity $(dz/dt)_i \approx 2\ 10^{-4}$ (m/s). These give $B = [(dz/dt)_i + \lambda_1(z_i - d/c)]/[(\lambda_1 - \lambda_2)\ \exp{-}(\lambda_2 t_i)]$ and $A = [(z_i - d/c) - B\ \exp{-}(\lambda_2 t_i)]/\exp{-}(\lambda_1 t_i)$. Replacing $A$ and $B$ into Eq. 6 yields,

$$z(t) = d/c + [(z_i - d/c) - ((dz/dt)_i + \lambda_1(z_i - d/c))/(\lambda_1 - \lambda_2)]\ \exp{-}(\lambda_1(t - t_i))$$
$$+ [((dz/dt)_i + \lambda_1(z_i - d/c))/(\lambda_1 - \lambda_2)]\ \exp{-}(\lambda_2(t - t_i)) \quad (7)$$
$$(dz/dt)(t) = -\lambda_1[(z_i - d/c) - ((dz/dt)_i + \lambda_1(z_i - d/c))/(\lambda_1 - \lambda_2)]\ \exp{-}(\lambda_1(t - t_i))$$
$$- \lambda_2[((dz/dt)_i + \lambda_1(z_i - d/c))/(\lambda_1 - \lambda_2)]\ \exp{-}(\lambda_2(t - t_i)) \quad (8)$$

We depict in Fig. 5a the sinking velocity $dz/dz$ versus depth $z$ using Eqs. 7 and 8 with the static friction coefficients $\mu_0$ and effective viscosity $\eta_{eff}$ derived from the steady sinking regime for $\Gamma = 1.08$. The solution describes correctly the sinking data at long time (linear dependence) as expected, but not at all the behaviour in the acceleration regime at short time. This is likely caused by the breakdown of the approximation on the homogeneous friction coefficients. To account for the spatial inhomogeneity of the liquefied layer by horizontal vibration, we propose an empirical depth-dependent viscous friction coefficient $\eta_{eff} = \eta_{max}\ [1 - \exp(-k_1 x)]$ with $k_1 \sim 15$ and $x = 0.01$ to 1 the dimensionless parameters corresponding to the sinking range. From this assumption we obtain an increase of the viscous coefficient by a factor of about 20 with $\eta_{min} = 4.7\ 10^3$ to $\eta_{max} = 8.7\ 10^4$ over the whole depth of sinking (Fig. 5b). With such depth varying $\eta_{eff}$, the computed solution with Eqs. 7 and 8 provides a more consistent description of the experimental observation (Fig. 5c).

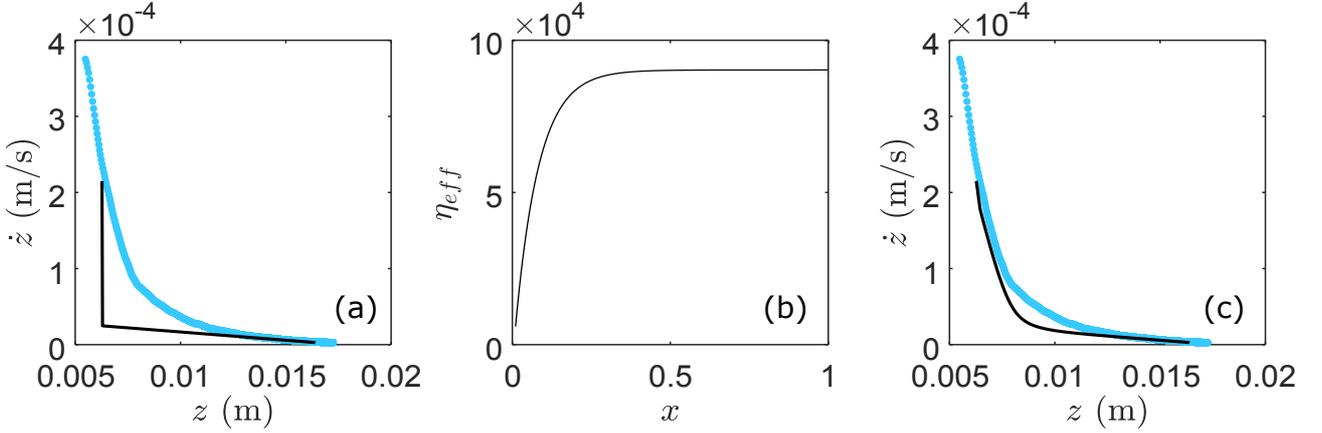

**Fig. 5** . (a) Comparision between model and experiment with homogeneous $\eta_{eff}$; (b) depth dependent $\eta_{eff}$; (c) comparision with inhomogeneous $\eta_{eff}$.

**References**

1. Liu, A.J. and Nagel, S.R. *Jamming and Rheology* (Taylor & Francis, New York, 2001)
2. van Hecke, M. Jamming of Soft Particles: Geometry, Mechanics, Scaling and Isostaticity. *J. Phys.: Condens. Matter* **80**, 033101 (2010)
3. Coussot, P. *Rheometry of pastes, suspensions and granular materials* (Wiley, New York, 2005)
4. Andreotti, B., Forterre, Y. and Pouliquen, O. *Granular Media: between fluid and solid* (Cambridge University Press, 2013)
5. Kamrin, K. and Goddard, J. Rheology of sedimenting particle pastes. Proc. R. Soc. A. 470, 2172 (2014)
6. Dedegil, M.Y. Drag coefficient and settling velocity of particles in non-Newtonian suspensions, *J. Fluids Eng.* **109**, 319 (1987)
7. Beris, A.N. Creeping motion of sphere through a Bingham plastic, *J. Fluid Mech.* **158**, 219 (1985)
8. Fall, A., de Cagny, H., Bonn, D., Ovarlez, G., Wandersman, E. et al. Rheology of sedimenting particle pastes. *J. Rheol.* **57**, 1237 (2013)
9. Boyer, F., Guazzelli, E., and Pouliquen, O. Unifying suspension and granular rheology. *Phys. Rev. Lett.* **107**, 188301 (2011)
10. H. M. Jaeger, C.H. Liu, and S. R. Nagel, T.A. Witten, Friction in granular flows. *Europhys. Lett.* **11**, 619 (1990)
11. GDR MiDi, On dense granular flows. *Euro. Phys. J. E* **14**, 341 (2004)

12. Courrech du Ponts, S., Gondret, P., Perrin, B. and Rabaud, M. Granular avalanches in fluids. *Phys. Rev. Lett.* **90**, 044301 (2003)
13. Cassar, C., Nicolas, M. and Pouliquen, O. Submarine granular flows down inclined planes. *Phys. Fluids* **17**, 103301 (2005)
14. Kamrin, K. and Koval, G. Nonlocal constitutive relation for steady granular flow. *Phys. Rev. Lett.* **108**, 178301 (2012)
15. Bouzid, M., Izzet, A., Trulsson, M., Clément, E., Claudin, Ph., and Andreotti, B. Non-local rheology in dense granular flows. *Eur. Phys. J. E* **38**, 125 (2015)
16. Nichol, K., Zanin, A., Bastien, R., Wandersman, E., and van Hecke, M. Flow-induced agitations create a granular fluid. *Phys. Rev. Lett.* **104**, 48 (2010)
17. Reddy, K., Forterre, Y., and Pouliquen, O. Evidence of mechanically activated processes in slow granular flows. *Phys. Rev. Lett.* **106**, 108301 (2011)
18. Jaeger, H.M., Liu, C.H., and Nagel, S.R. Relaxation at the angle of repose. *Phys. Rev. Lett.* **62**, 40 (1989)
19. D'Anna, G., Mayer, P., Barrat, A., Loreto, V., and Nori, F. Observing brownian motion in vibration-fluidized granular matter. *Nature* **424**, 909 (2003)
20. Metcalfe, G., Tennakoon, S., Kondic, L., Schaeffer, D., and Behringer, R. Granular friction, Coulomb failure, and the fluid-solid transition for horizontally shaken granular materials. *Phys. Rev.* **65**, 031302 (2002)
21. Aumaitre, S., Puls, C., McElwaine, J., and Gollub, J. Comparing flow thresholds and dynamics for oscillating and inclined granular layers. *Phys. Rev. E* **75**, 061307 (2007)
22. Marchal, P., Smirani, N., and Choplin, L. Rheology of dense-phase vibrated powders and molecular analogies *J. Rheol.* **53**, 1 (2009)
23. Dijksman, J., Wortel, G., van Dellen, L., Dauchot, O., and van Hecke, M. Jamming, yielding, and rheology of weakly vibrated granular media. *Phys. Rev. Lett.* **107**, 108303 (2011)
24. Johnson, P. and Jia, X. Nonlinear dynamics, granular media and dynamic earthquake triggering. *Nature* **437**, 871 (2005)
25. Lastakowski, H., Géminard, J.-C., and Vidal, V. Granular friction: Triggering large events with small vibrations. *Sci. Rep.* **5**, 13455 (2015)
26. Hanotin, C., Kiesgen de Richter, S., Marchal, P., Michot, L. and Baravian, C. Vibration-induced liquefaction of granular suspensions. *Phys. Rev. Lett.* **108**, 198301 (2012)
27. Gaudel, N., Kiesgen de Richter, S., Louvet, N., Jenny, M., and Skali-Lami, S. Bulk and local rheology in a dense and vibrated granular suspension. *Phys. Rev. E* **96**, 062905 (2017)
28. Khaldoun, A., Eiser, E., Wegdam, G.H., and Bonn, D. Rheology: liquefaction of quicksand under stres. *Nature* **437**, 635 (2005)
29. Clément, C., Toussaint, R., Stojanova, M., and Aharonov, E. Sinking during earthquakes: Critical acceleration criteria control drained soil liquefaction. *Phys. Rev. E* **97**, 022905 (2018)
30. Jia, X., Brunet, T., and Laurent, J. Elastic weakening of a dense granular pack by acoustic fluidization: slipping, compaction, and aging. *Phys. Rev. E* **84**, 020301 (R) (2011)
31. Brum, J., Gennisson, J.-L., Fink, M., Tourin, A., and Jia, X. Drastic slowdown of shear wave in unjammed granular materials (to be submitted)
32. Harisch, R., Darnige, T., Kolb, E., Clément, E. Intruder mobility in a vibrated granular packing. *Europhys. Lett.* **96**, 54003 (2011)
33. Bureau, L., Caroli, C., and Baumberger, T. Elasticity and onset of frictional dissipation at a non-sliding multi-contact interface. *Proc. R. Soc. Lond. A.* **459**, 2787 (2003)
34. Léopoldès, J., Conrad, G., and Jia, X. Onset of sliding in amorphous films triggered by high-frequency oscillatory shear. *Phys. Rev. Lett.* **110**, 248301 (2013)
35. DeGiuli, E. and Wyart, M. Friction law and hysteresis in granular materials *PNAS* **114**, 9284 (2017)
36. Goddard, J. (private communications)
37. Risten, H. *The Fokker-Planck equation: methods of solution and applications* (Springer, Berlin 1996)
38. Makse, H. and Kurchan, J. Testing the thermodynamic approach to granular matter with a numerical model of a decisive experiment. *Nature* **415**, 614 (2002)

**Acknowledgment**

We thank J. Goddard for the critical reading of the manuscript and references on the ball falling in Bingham fluids, J.-L. Gennisson and A. Trabattoni for helpful assistances. This work was supported by French LABEX WIFI under references ANR-10-LABX-24 and ANR-10- IDEX-0001-02 PSL.

# Supplementary information

The list of supplementary information files is provided below with a description of their content :

Supplementary Movie S1 (downloaded at https://dl.espci.fr/ticket/26a780341030a2c9338f50f7544aefd5):
Example video of a steel ball ($R$ = 10 mm) sinking in a dry granular media (packing of glass beads with $d$ = 100 μm) under vibration of $\Gamma$ = 0.5. The movie is taken from the top and looking down on the ball at a frame rate of 30 Hz. The movie is sped up 8 times. Light scintillation reveals the motion of glass beads (translation/rotation) induced by the external (horizontal) vibration far away from the ball (also visible without the sinking ball).

Supplementary Movie S2 (downloaded at https://dl.espci.fr/ticket/b514279778f05e320d9818a5b8131760):
Example video of a steel ball ($R$ = 10 mm) sinking in a saturated granular suspension (glass beads with d= 100 μm in water) under vibration of $\Gamma$ = 0.5. The movie is taken from the top and looking down on the ball, at a frame rate of 30 Hz. The movie is sped up 8 times. In contraste to the dry case, we do not observe motion of glass beads induced far away from the intruder.

Supplementary Figure S3 : Position of the ball as a function of time obtained in dry granular packinkgs and saturated granular suspensions for different vibration intensities. The data was obtained from movies such as shown in S1and S2 as follows. In each frame we determined the observed radius $r$ by finding the perimeter of the ball. The position of the ball in the frame was then calculated via $z = (R^2 - r^2)^{1/2}$.

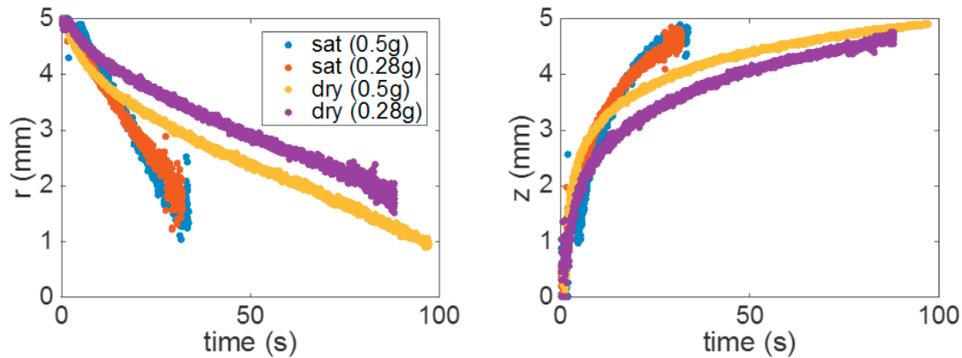

Figure S3